# Effect of impurities on growth and morphology of cementite nanowires

D.W. Boukhvalov[1,2,*], M. I. Katsnelson[3], Yu. N. Gornostyrev[2,4]

[1]School of Computational Sciences, Korea Institute for Advanced Study (KIAS) Hoegiro 87, Dongdaemun-Gu, Seoul, 130-722, Korean Republic

[2]Institute of Quantum Materials Science, Ekaterinburg 620175, Russia

[3]Radboud University Nijmegen, Institute for Molecules and Materials Heyendaalseweg 135, 6525AJ, Nijmegen, the Netherlands

[4]Institute of Metal Physics, Russian Academy of Sciences-Ural Division, 620041 Ekaterinburg, Russia

*Effects of doping on morphology of iron carbide (cementite) nanowires have been explored by first principle electronic structure calculations. We examined the role of several realistic impurities (Si, Mn, V, P and S) in the formation energies of the cementite nanowires with different sizes and morphologies. It is shown that the presence of the impurities decreases the formation energy and can switch the preferable axis of the cementite nanowire growth. The conditions of the formation and decomposition of cementite nanowires in steels are also discussed.*

E-mail: danil@kias.re.kr

# 1. Introduction

Fabrication of the nanowires and nanorods with a given morphology is one of the most explored subjects in modern nanoscience [1-21]. Whereas semiconductor [1-7], elemental-metal [8-11] and metal-oxide [12-15] nanowires were discussed in plenty of works, the formation of metal-carbide nanostructures was reported only for the several metals (Al, Ti and Mo) [16-19]. Recent experimental works [5-7] for the case of semiconductor nanowires demonstrate that impurities can be catalysts of the nanowire formation and determine its morphology. These results show that doping can play the role not only in the conditions of production of nano-objects but also in their morphology and phase transitions there. In previous theoretical works the segregation of various impurities near the surface of semiconducting nanowires has been explored [20-25]. The changes in the electronic structure caused by the presence and segregation of impurities were considered. However, neither influence of impurities to the formation energies nor possible changes in the morphology were discussed. It is important therefore to examine the role of doping for the morphology of realistic nanowires.

Iron carbide ($Fe_3C$, or cementite) nanostructures have been obtained as secondary phases in the processes of fabrication of iron-filled carbon nanotubes [26, 27]. Detection of the cementite nanowires in the ancient ($12^{th}$ century AD) sabers [28] made from the famous Damascus steel (further DS) [29] arose the question about necessary conditions for the growth of metal-based nanowires. So, studies of the physical mechanisms of the formation of cementite nanowires have both fundamental interest and importance for the whole nanofabrication area. The experimental results [30-34] give evidences of an important role of impurities for unusual structure and properties of DS. It requires a detailed survey of the role of realistic impurities in the formation and morphology of the freestanding cementite nanowires.

In our work we study the energetics of the pure and doped cementite nanowires grown along three possible directions and discuss the possible ways of decomposition of the cementite nanowires with formation of different carbon-based nanostructures [35, 36]. We show that different impurities result in the growth of the cementite nanowires along different axes.

## 2. Computational method and model

The modeling is carried out using the density functional theory (DFT) implemented in the pseudopotential SIESTA code [37]. All calculations are done in the generalized gradient approximation (GGA) [38] with spin polarization. Full optimization of the atomic positions was performed. During the optimization, the ion cores are described by norm-conserving non-relativistic pseudo-potentials [39] with cut off radii about 2.25 a. u. for *3d* elements and about 1.25 a.u. for light elements used in calculations, and the wavefunctions are expanded with a double-ζ plus polarization basis of localized orbitals for carbon and oxygen, and a double-ζ basis for hydrogen. Optimization of the force and total energy was performed with an accuracy of 0.04 eV/Å and 1 meV, respectively. All calculations were carried out with an energy mesh cut-off of 360 Ry and a k-point mesh of 4×4×4 in the Mokhorst-Park scheme [40]. These technical parameters are chosen similar to those in our previous work on iron-carbon systems [41]. We define the formation energy of the cementite nanowires as $E_{form} = E_{nanowire} - E_{bulk}$, where $E_{bulk}$ is the total energy of cementite supercell in the bulk in the same magnetic configuration. For the case of cementite with impurities $E_{bulk}$ is the total energy for the supercell of cementite with the same concentration and disposition of the impurities as in the nanowires. For the simulation of nanowires we use cementite supercell (see Fig. 1) within the periodical boundary conditions along selected crystallographic axises and separated by 5 nm of empty space between nanowires.

## 3. Results and discussions

In the present work we consider the effect of different impurities on the morphology of cementite nanowires. We have performed our modeling for both pure and doped nanowires oriented along different crystallographic axes. Mn and Si were chosen as the most common impurities (of course, apart from carbon) in ordinary steels, V is also used in the high-strength low-alloy steels [29] and P, S and V were experimentally found in DS [27, 28, 34]. The calculations were performed first for the ferromagnetic ground state of cementite. However, the morphology of cementite in steels at high enough temperature (in particular, above the Curie temperature) is also of great interest. To study the effects of changes in magnetic state induced by the temperature we have made also the calculations for disordered magnetic configurations [42, 43]. More specifically, the calculations for three different randomly chosen spin configurations with the total magnetic moments equal to zero, which models the high-temperature paramagnetic state, have been performed for the bulk cementite and the nanowires. During the optimization of atomic structure we keep the chosen magnetic configuration. In all studied disordered configurations we have found a decrease of formation energies of cementite nanowires. For the smallest (1.24 nm) width of the nanowire oriented along $c$ axis it changes from 5.10 to 4.95 eV/$Fe_3C$ for the pure case, from 4.95 to 4.70 eV/$Fe_3C$ for P-doped and from 5.21 to 5.13 eV/$Fe_3C$ for Mn-doped cases, in comparison with the ferromagnetic states. Thus, the degree of magnetic order of cementite plays an insignificant role in the energetics of the formation of pure and doped cementite nanowires.

In contrast to elemental iron, the crystal structure of cementite is strictly anisotropic and can be described as a layered system with alternation of two iron and one carbon layers perpendicular to the crystallographic $b$ axis (see Fig. 1). The axises in the cementite are chosen according to the standard denomination of the axises in cementite lattice used in the previous works [28]. This special crystal structure determines the anisotropy of lattice distortions in the presence of impurities. Our calculations demonstrate that all studied in this work impurities (Mn, V, Si, P, S) produce an expansion of cementite along $a$ and $c$ axises and an insignificant compression along $b$ axis. Substitution of iron atom by silicon or manganese provides a smaller lattice expansion than the substitution of carbon atom by phosphorus or sulfur (0.5 and 2% for Mn and S, respectively). This tendencies in cementite lattice expansion is due to a big difference between carbon and phosphorus ionic radii (0.15 vs. 0.37 Å), so the difference of iron and manganese ionic radii being much smaller (both are about 0.6 Å). The insignificant changes of structural anisotropy for bulk cementite discussed above can be more important in the nanophases.

Further modeling requires a specification of the model for the chemical composition of cementite nanowires. First of all, we studied the disposition of impurities inside the nanowires. We calculated the total energies of cementite nanowires of the width of 4.44 nm with Mn or P impurities situated at different distances from the center of the wire. For both types of impurities the substitution of the surface host atom by the impurity atom (see Fig. 1b-c) turned out to be more favorable than the substitution of the host atom in the center of the nanowire. For manganese and phosphorus this difference is 0.48 eV/atom and 0.72 eV/atom, respectively. The calculated energy difference between these two types of impurities is caused by the deeper relaxation near the surface of nanowires in contrast to the bulk region. This result is similar to the previously obtained computational results for the impurities segregation in semiconductor nanowires [20-25].

The second step of the model definition is the evaluation of the length and the width of the supercell used for the calculations of doped cementite nanowires. We examined the role of width and length for manganese and phosphorous as typical impurities for the substitution of iron (Mn) and carbon (P) in cementite. For computational simplicity we used for this check the cementite nanowire of the minimal width only (0.6 nm) shown in Fig. 1b. We increased the length of the supercell, which corresponds to the raise of the distance between the impurities along the nanowire. The computational results (Fig. 2a) suggest that the distances between impurities along the nanowire are insignificant. For the case of impurities separated by the distances more than 1 nm the effect is really small, due to a decay of the lattice distortions. In contrast with the length of nanowires, the increase of their widths provides an exponential decrease of the formation energy for pure and doped cementite nanowires (Fig. 2b). At further increase of the nanowire width the properties of nanosystems approach those of the bulk, the formation energies decreasing. This situation is opposite to the increasing of the distance between impurities along nanowires because there relatively weak bonds between layers makes the properties of supercell with minimal length approximate to the properties of bulk cementite. Further calculation of dependence of formation energies from width of the cementite nanowires (see Fig. 3) demonstrate similar tendencies for the decrease of formation energies with the increase of the radii of the cementite nanowires for the both types of impurities (V, Mn, Si as substitutes of iron vs. P, S as substitutes of carbon).

Next, we consider the results for the nanowires of different width (1.24, 2.60 and 4.44 nm) with the minimal length of the supercell. The smallest value corresponds to the minimal width of the cementite unit cell and the bigger one is close to the experimentally observed thickness of these nanowires [22]. First of all, it is worthwhile to note that, independently on the type of impurities and the width of nanowires, the formation energy for the nanowires grown along *a* axis is higher than that for other directions, the difference is about 0.4 eV per $Fe_3C$ unit. Further we will discuss only the results for the nanowires oriented along *b* and *c* axises. The computational results are presented on Fig. 3. One can conclude from the obtained data that vanadium plays no role in the morphology of cementite nanowires. Manganese and silicon strongly suppress the formation of cementite nanowires oriented along *b* axis. For pure cementite nanowires with a larger width the tendency of growth along the same axis was also observed. The presence of Mn and Si impurities only decreases the formation energy of nanowires, due to a decrease of the corresponding surface energy.

For the cases of phosphorus and sulfur the aforementioned reduction of the formation energy have been obtained. However, it is worthwhile to note that the reduction of the formation energy for P and S doping is stronger than for Mn and Si and for the nanowires of realistic sizes this decrease is about 1/6 and the resulting formation energy of the doped nanowire is less than 1.5 eV (Fig. 3c). We have also examine the role of combination of similar (Mn+Si and P+S) and different (P+Si) impurities and found that the combination of manganese or phosphorous with silicon impurities does not change significantly formation energies of the cementite nanowires in contrast to the combination of phosphorous and sulfur which is significantly decrease the formation energy for the case of the nanowire grown along $c$ axis. More significant changes of the structural anisotropy of the bulk cementite provide a switch of the preferable axis from $b$ to $c$.

Since the magnetic order does not effect qualitatively on the energetics of nanowires, as was discussed above, these results seem to be feasible for the temperatures above Curie temperature as well. As a result, the changes in the morphology of doped cementite nanowires considered here should take place also for the realistic conditions of metallurgical processes. Ancient DS samples contain an enormous amount of phosphorous and sulfur and a rather low amount of manganese and silicon. Interestingly, the cementite nanowires oriented along $c$ axis were found in P and S-rich samples of DS [31, 34]. Our results explain this observation.

For the temperatures above 750 °C the decomposition of bulk cementite into α-iron and Kish-graphite is observed. This process can be described as a removal of iron layers from the spaces between carbon layers in bulk cementite keeping the layered structure of carbon that leads to the formation of graphite and iron [44]. In the case of decomposition of the cementite nanowires different carbon nanostructures could appear. For the nanowires oriented along $b$ axis with carbon layers perpendicular to the main axis the formation of nano-graphites can be realized. This case is similar to the ordinary decomposition of bulk cenemtite with Kish–graphite formation.  In the opposite case (nanowires grown along $c$ axis) multilayer graphene nanoribbons of few nanometers widths can be probably obtained. We can speculate that, according to the recent experiments for multiwall carbon nanotubes unzipping to graphene nanoribbons [45, 46], a reverse process can be realize at high temperatures and oxygen-free conditions as it was discussed in Ref. [47]. This process may cause an intriguing appearance of multiwall carbon nanotubes in the ancient DS sabers [36] which can be formed from experimentally observed there cementite nanowires oriented along c axis [28]. Thus, we can conclude that the presence of impurities can play a role not only for the formation and morphology of nano-objects but, in the case of cementite nanowires, it could be important for the formation of other nanostructures appeared after cementite decomposition by annealing in the metallurgical processes.

## 4. Conclusions

In summary, we can conclude that the performed DFT modeling demonstrates an important role of impurities in the growth and morphology of the cementite nanowires. All impurities considered here decrease the formation energy of cementite nanowires and make cementite nanowires more robust in realistic environment. Manganese and silicon impurities, common in usual steels, cause the growth of these nanowires along *b* axis. Contrary, the presence of phosphorus and sulfur lead to the most significant decrease of the formation energy of cementite nanowires and their growth along *c* axis. These results may be relevant to explain the formation of cementite nanowires oriented along c axis in ancient samples of Damascus steel, which is usually phosphorous and sulfur rich. The switch of the grown axis of cementite nanowires by the different dopants found in our modeling corresponds to the different structural distortions of the bulk cementite by the same impurities. Doping as a source of enhancement or suppression of anisotropy of host materials can play an important role for the design of novel nanomaterials.

**Acknowledgements** DWB thank Korea Institute for Advanced Study for providing computing resources (KIAS) Center for Advanced Computation Linux Cluster System) for this work. MIK acknowledges support from the "Stichting voor Fundamenteel Onderzoek der Materie (FOM)," which is financially supported by the "Nederlandse Organisatie voor Wetenschappelijk Onderzoek (NWO)."

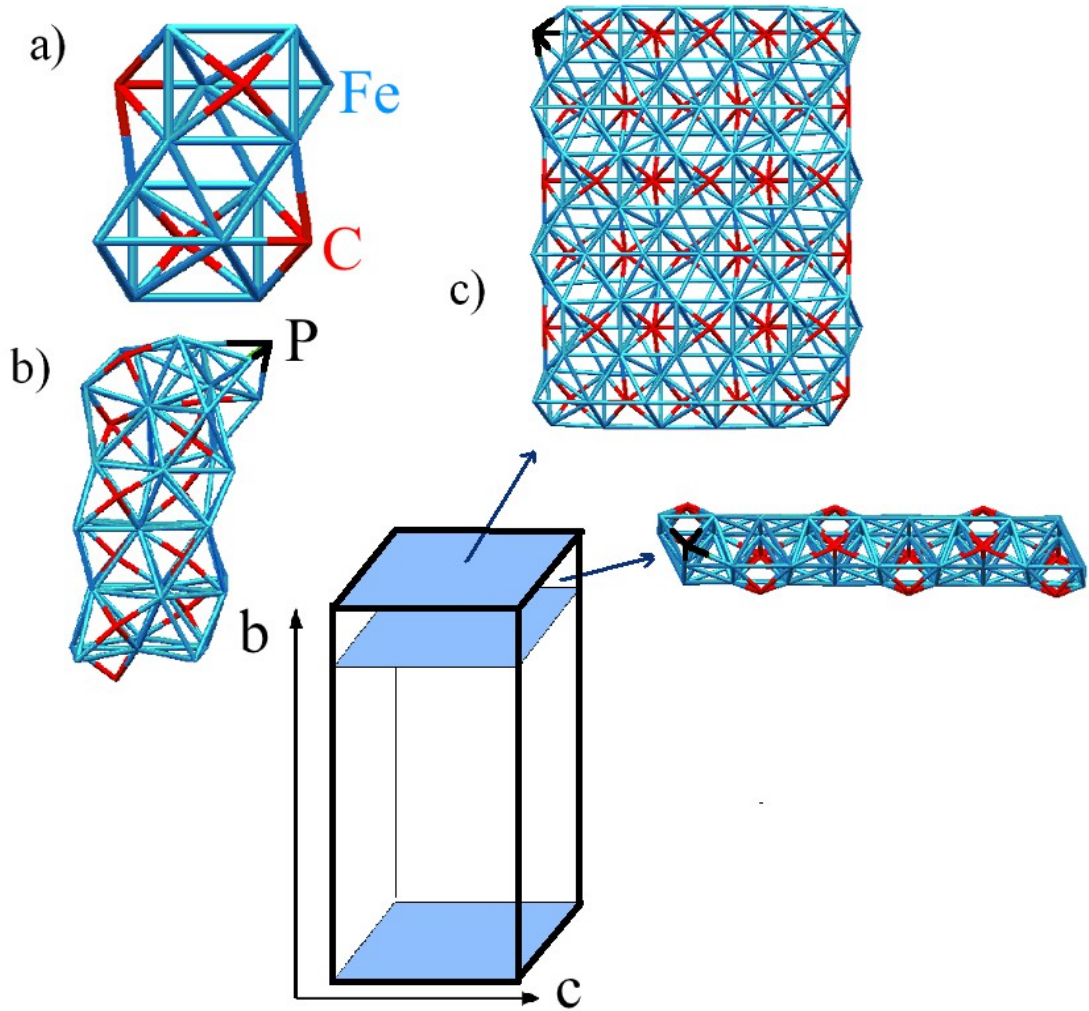

**Figure 1**. Optimized crystal structure for cementite unit cell (a); supercell of cementite nanowire with the minimal width (1a×3b×1c) doped by P (b); side and top view of supercell with the minimal length and with the width 4.44 nm (3a×1b×3c) also doped by P (c). Iron atoms are shown by blue, carbon by red, and phosphorus by black.

**Figure 2** Formation energy of pure (dashed green line), Mn-doped (solid red line) and P-

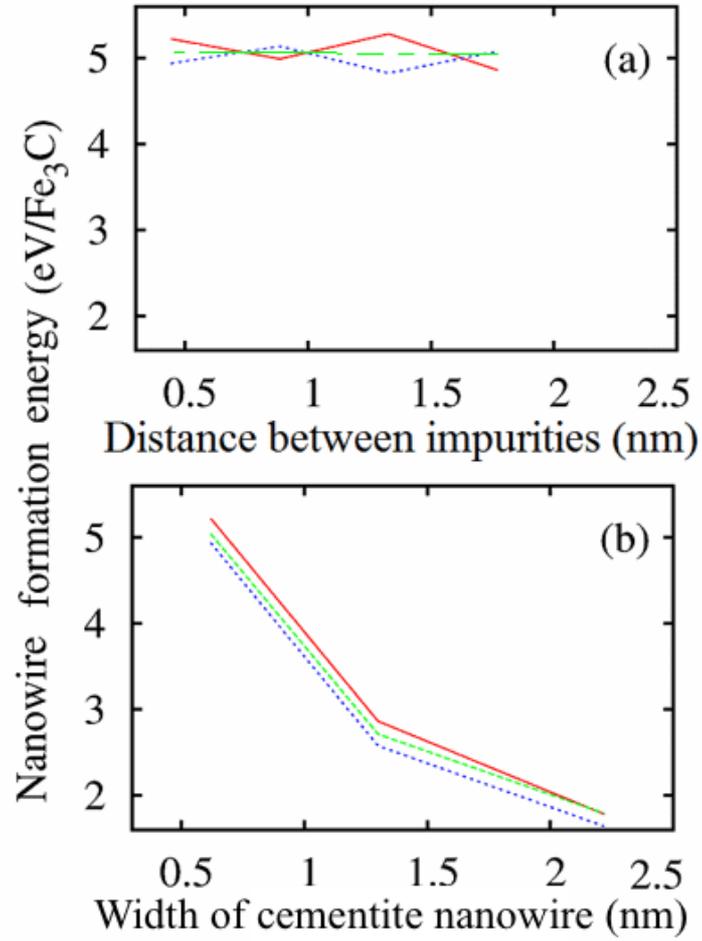

doped (doted blue line) cementite nanowires as a function of the distance between impurities in nanowire of minimal width (a) and nanowire width for the supercell of minimal length (b).

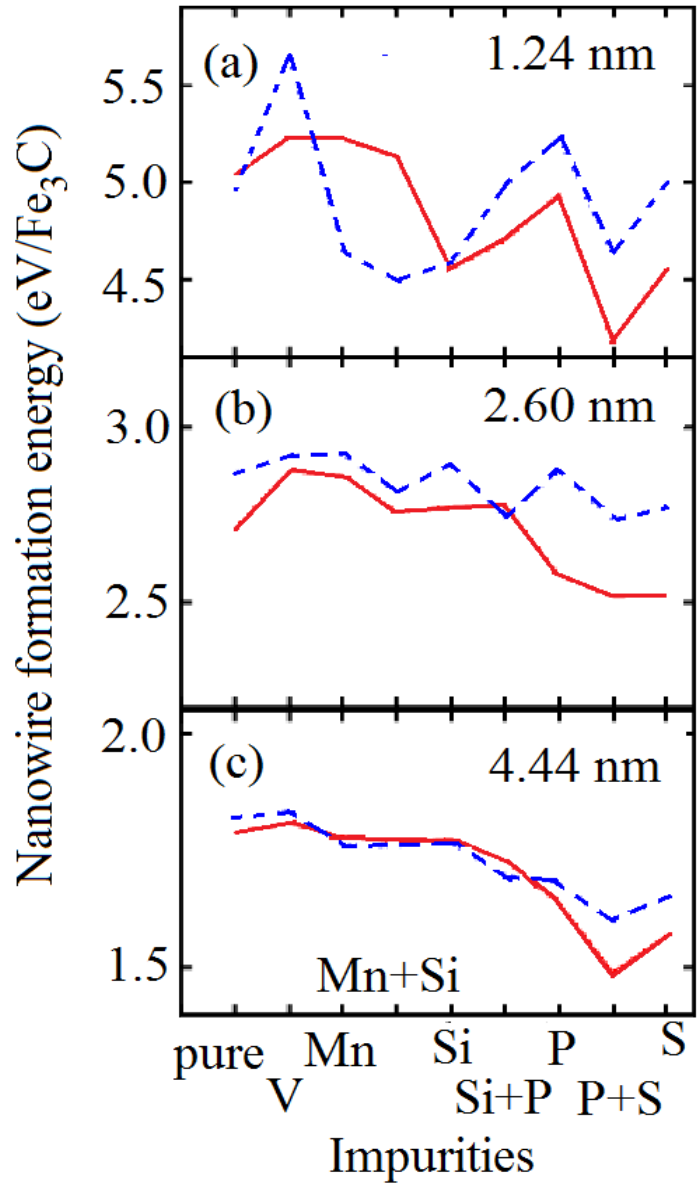

**Figure 3** Formation energy of cementite nanowires dependent on the type of impurities for different nanowire morphology (growing along *c* axis are shown by solid red line, along *b* axis by doted blue line) for the case of different width.